\documentclass[12pt]{article}
\usepackage{epsfig}
\usepackage{a4wide}
\newcommand\keV{{\rm\,keV}}
\newcommand\MeV{{\rm\,MeV}}
\newcommand\GeV{{\rm\,GeV}}
\newcommand\imag{{\rm Im\,}}
\newcommand\pfrac[2]{\left(\frac{#1}{#2}\right)}

\begin{document}
\thispagestyle{empty}
\begin{flushright}
MZ-TH/03-14\\
hep-ph/0309226\\
September 2003\\
\end{flushright}
\vspace{0.5cm}
\begin{center}
{\Large\bf QCD improved determination}\\[.3cm]
{\Large\bf of the hadronic contribution to the}\\[.3cm]
{\Large\bf anomalous magnetic moment of the muon}\\[1.3cm]
{\large S.~Groote, J.G.~K\"orner and J.~Maul}\\[.7cm]
Institut f\"ur Physik, Johannes-Gutenberg-Universit\"at,\\[.2cm]
Staudinger Weg 7, 55099 Mainz, Germany
\end{center}

\begin{abstract}
We present the results of a new evaluation of the anomalous magnetic moment
$a_\mu=(g_\mu-2)/2$ of the muon where the role of input data needed in the
evaluation is lowered in the interval between $1.2$ and $3.0\GeV$ below charm
threshold. This is achieved by decreasing the size of the weight function in
the dispersion integral over the experimental ratio $R(s)$ by subtracting a
polynomial from the weight function which mimics its energy dependence in that
given energy interval. In order to compensate for this subtraction, the same
polynomial weight integral is added again but is now evaluated on a circular
contour in the complex plane using QCD and global duality. For the hadronic
contribution to the shift in the anomalous magnetic moment of the muon we
obtain $a_\mu^{\rm had}({\rm LO})=(701.3\pm 6.4)\times 10^{-10}$ at leading
order in the electromagnetic coupling. In addition, using the same procedure,
we recalculate the next-to-leading contribution
$a_\mu^{\rm had}({\rm NLO})=(-10.3\pm0.2)\times 10^{-10}$. Including QED,
electroweak, and light-by-light contribution, we obtain a value
$a_\mu=(11\,659\,185.6\pm6.4_{\rm had}\pm3.5_{\rm LBL}\pm0.4_{\rm QED+EW})
\times 10^{-10}$.
\end{abstract}

\newpage

\section{Introduction}
The anomalous magnetic moment $a_\mu$ of the muon is one of a few physical
parameters which can be determined with high precision and therefore can serve
as precision test for the Standard Model of elementary particle physics. In
this paper we join the attempts to determine the hadronic contribution of the
anomalous magnetic moment of the muon with a better accuracy (for an overview
over the status of calculations of different collaborations two years ago see
e.g.\ Ref.~\cite{CzarneckiMarciano,MarcianoRoberts}). While the dominant QED
contribution $a_\mu^{\rm QED}=(11\,658\,470.6\pm 0.3)\times 10^{-10}$ and the
weak contribution $a_\mu^{\rm weak}=(15.4\pm 0.1\pm 0.2)\times 10^{-10}$ are
known with very good accuracy (see e.g.\
Refs.~\cite{Hughes:fp,Czarnecki:1998nd,Czarnecki:2002nt} and references
therein), the main uncertainty is given by the hadronic contribution. As an
example for the actual calculations of the leading order hadronic contribution
we cite the value
$a_\mu^{\rm had}=(683.1\pm 5.9 \pm 2.0)\times 10^{-10}$~\cite{Hagiwara}.

The calculation of the hadronic contribution to the anomalous magnetic moment
of the muon mostly relies on experimental data in the $e^+e^-$ channel. In
principle one can also make use of data from $\tau$-decay~\cite{Aleph}.
However, the inclusion of $\tau$-decay data introduces systematic
uncertainties originating from isospin symmetry breaking effects which are
difficult to estimate~\cite{Hoecker}. We have therefore decided to only
include $e^+e^-$ data in our analysis.

There are different concepts for using the $e^+e^-$ data sets, either to its
full extent~\cite{EidelmanJegerlehner} or in part by dividing the energy
 range into resonance regions close to the pair production
thresholds where the experimental values are taken and those regions far from
thresholds where perturbation theory is assumed to be valid. In this
paper we present an approach which has been successfully applied for the
determination of the running fine structure constant at the $Z^0$ boson
resonance, $\alpha(M_Z^2)$~\cite{GKSN}.

The subject of a precision determination of the anomalous magnetic moment
became rather important again as a precision measurement for the positive muon
could be accomplished at the Brookhaven National Laboratory
(BNL)~\cite{BNL:2001}. The value
\begin{equation}
a_\mu^{\rm exp}=(11\,659\,202.3\pm 15.1)\times 10^{-10}
\end{equation}
cited in Ref.~\cite{BNL:2001} showed a deviation from the Standard Model
prediction at that time,
\begin{equation}
a_\mu^{\rm SM}=(11\,659\,159.7\pm 6.7)\times 10^{-10}
\end{equation}
as weighted average over the calculation results of different
collaborations~\cite{MarcianoRoberts}. The deviation was $2.6$ standard
deviations which seemed to have opened the window for new physics. Many such
suggestions were published, including concepts of supersymmetry,
leptoquarks, lepton number violating models, technicolor models, string theory
concepts, extra dimensions and so on. The discrepancy between measurement and
the Standard Model prediction became smaller, though, when a sign error was
discovered in the theoretical calculation of the light-by-light
contribution~\cite{Hayakawa:2001bb}. Therefore, it is worth still to analyze
the situation of the Standard Model prediction thoroughly, giving special
emphasis on the error estimate, in order to compare it with the experimental
world average
\begin{equation}
a_\mu^{\rm exp}=(11\,659\,203\pm 8)\times 10^{-10}
\end{equation}
which is dominated by the measurement of Ref.~\cite{BNL:2002}, as we will do
in this paper.

\section{Theoretical background}
\begin{figure}\begin{center}
\epsfig{figure=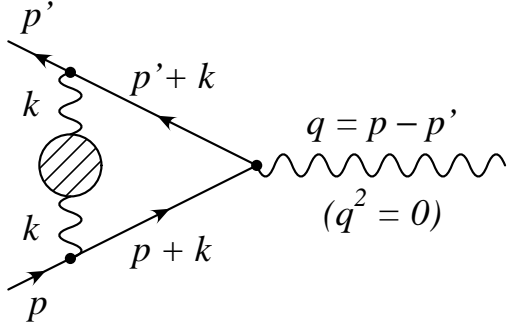,scale=0.5}
\end{center}
\caption{\label{fig1}$O(\alpha^2)$ correction to the hadronic part of $a_\mu$.
The shaded bubble represents the current contributions to the vector part of
the hadronic vacuum polarisation given in Refs.~\cite{Harlander,Gorishnii}.}
\end{figure}

The leading hadronic contribution to the anomalous magnetic moment of the muon
$a_\mu^{\rm had}$ can be extracted from the $O(\alpha^2)$ vertex correction
shown in Fig.~\ref{fig1}.
The analytic expression for $a_\mu^{\rm had}$ from this diagram is a weighted
dispersion integral over the imaginary part of the vector part of the hadronic
vacuum polarisation,
\begin{equation}\label{disper}
a^{\rm had}_\mu=\frac{4\alpha^2}\pi\int_{4m_{\pi}^2}^\infty
  \imag\Pi^{\rm had}(s)\frac{K(s)}sds
\end{equation}
with the QED kernel
\begin{eqnarray}
K(s)&=&\ln(1+x)\frac{(1+x^2)(1+x)^2}{x^2}
  +\frac{x^2(1+x)}{1-x}\ln(x)\nonumber\\&&
  +\frac{x^2}2(2-x^2)+\frac{(1+x^2)(1+x)^2}{x^2}\left(-x+\frac{x^2}2\right).
\end{eqnarray}
As usual we make use of the kinematic variables
\begin{equation}
x(s)=\frac{\beta(s)-1}{\beta(s)+1},\qquad\beta(s)=\sqrt{1-\frac{4m_\mu^2}s}.
\end{equation}

$\imag\Pi^{\rm had}(s)$ is calculated from the hadronic current-current
correlator,
\begin{equation}
\Pi^{\rm had}_{\mu\nu}(q^2)=i\int e^{iqx}
\langle0|T\;j_\mu(x)j_{\nu}(x')|0\rangle d^4x
  =\Pi^{\rm had}(q^2)\left(q^2g_{\mu\nu}-q_\mu q_\nu\right)
\end{equation}

Higher perturbative QCD corrections are well-known up to $O(\alpha_s^2)$ with
$O(m_q^{12}/q^{12})$ quark mass corrections which are supplemented by massless
terms up to $O(\alpha_s^4)$~\cite{Harlander}. This perturbative part is
denoted by $\Pi^{\rm P}(q^2)$. For brevity we only cite the first few terms of
this expansion in $\alpha_s$ and $m_q^2/q^2$,
\begin{equation}
\Pi^{\rm P}(q^2)=\frac3{16\pi^2}\sum_{i=1}^{n_f}Q_i^2\Bigg[
  \frac{20}9+\frac43L+C_F\left(\frac{55}{12}-4\zeta(3)+L\right)
  \frac{\alpha_s}\pi+O(\alpha_s^2,m_q^2/q^2)\Bigg]
\end{equation}
with $L=\ln(\mu^2/q^2)$. In our analysis we use the full
$O(\alpha_s^2,m_q^{12}/q^{12})$ expression given in Ref.~\cite{Harlander}. The
number of active flavours is denoted by $n_f$ which changes according to the
energy interval under consideration. For the zeroth order term in the
$m_q^2/q^2$ expansion we have included the known higher order terms in
$\alpha_s$,
\begin{equation}
\frac3{16\pi^2}\sum_{i=1}^{n_f}Q_i^2\Bigg[\left(c_3+k_2L
  +\frac12(k_0\beta_1+2k_1\beta_0)L^2+\frac13k_0\beta_0^2L^3\right)
  \left(\frac{\alpha_s}\pi\right)^3+O(\alpha_s^4)\Bigg]
\end{equation}
with $k_0=1$, $k_1=1.63982$ and $k_2=6.37101$. We have denoted the yet unknown
constant term in the four-loop contribution by $c_3$ which, however, does not
contribute to our calculations since it has no absorption part.

For the nonperturbative part the operator product expansion leads
to~\cite{Gorishnii}
\begin{eqnarray}
\lefteqn{\Pi^{\rm NP}(q^2)\ =\ \frac1{18(q^2)^2}
  \left(1+\frac{7\alpha_s}{6\pi}\right)
  \langle\frac{\alpha_s}{\pi}G^2\rangle}\nonumber\\&&
  +\frac8{9(q^2)^2}\left(1+\frac{\alpha_s}{4\pi}C_F+\ldots\ \right)
  \langle m_u\bar uu\rangle
  +\frac2{9(q^2)^2}\left(1+\frac{\alpha_s}{4\pi}C_F+\ldots\ \right)
  \langle m_d\bar dd\rangle\nonumber\\&&
  +\frac2{9(q^2)^2}\left(1+\frac{\alpha_s}{4\pi}C_F
  +(5.8+0.92L)\frac{\alpha_s^2}{\pi^2}\right)
  \langle m_s\bar ss\rangle\nonumber\\&&
  +\frac{\alpha_s^2}{9\pi^2(q^2)^2}(0.6+0.333L)
  \langle m_u\bar uu+m_d\bar dd\rangle\nonumber\\&&
  -\frac{C_Am_s^4}{36\pi^2(q^2)^2}
  \left(1+2L+(0.7+7.333L+4L^2)\frac{\alpha_s}{\pi}\right)
  -\frac{448\pi}{243(q^2)^3}\alpha_s|\langle\bar qq\rangle|^2\nonumber\\&&
  +\frac1{(q^2)^4}(0.48\GeV)^8-\frac1{(q^2)^5}(0.3\GeV)^{10}+O((q^2)^{-6})
\end{eqnarray}
where $C_F=4/3$, $C_A=3$, $T_F=1/2$ are $SU(3)$ colour factors. We will use
these results for the evaluation of the theoretical contributions.

\section{Experimental contributions}
The evaluation of the integral involves experimental data via the optical
theorem which connects the imaginary part of the vacuum polarisation with the
ratio
\begin{equation}
R(s)=\frac{\sigma(e^+e^-\rightarrow{\rm hadrons})}{\sigma(e^+e^-\rightarrow
  \mu^+\mu^-)}=12\pi\imag\Pi^{\rm had}(s).
\end{equation}
Especially in the $\MeV$ and the low $\GeV$ energy range and in regions of
resonances where perturbative QCD cannot be applied, data sets taken from
experimental measurements are crucial. For our evaluation we have used the
combined $e^+e^-$ data sets from Ref.~\cite{EidelmanJegerlehner,BES} and
complement them in the dominant low energy range from $610\MeV$ to $961\MeV$
by recent two pion data~\cite{Akhmetshin}. By comparing with the new data for
three and four pion decays~\cite{Achasov,Novosib} one ensures that their
influence for the sub $\GeV$-range where the two pion data were recorded is
neglible.

All data sets are combined by weighting their relative contribution to the
total experimental value with the corresponding statistic and systematic
errors.

\section{Introduction of the method}
Global duality states that QCD can be used in weighted integrals over a
spectral function if the spectral function is multiplied by polynomial
functions. However, this does not work if the polynomial function is replaced
by a singular function such as the weight $K(s)/s$ in the present case.
Nevertheless, local duality is expected to hold for large values of $s$
far from resonances and threshold regions, i.e.\
$\imag\Pi^{\rm had}(s)\simeq\imag\Pi^{\rm QCD}(s)$.

In our approach we attempt to reduce the influence of experimental $R(s)$ data
in regions where the data has large uncertainties. Specifically, this holds
for the interval between $1.2$ and $3.0\GeV$. The essence of our method is to
diminish the magnitude of the weight function by subtracting a polynomial
function which mimics the weight function at those energies. In order to
compensate for this subtraction, the same polynomial function is added again,
but now its contribution is evaluated by using global duality on a circular
contour in the complex plane, according to
\begin{eqnarray}
\lefteqn{\frac{\alpha^2}{3\pi^2}\int_{s_1}^{s_2}R(s)\frac{K(s)}{s}ds
\ =\ \frac{\alpha^2}{3\pi^2}\int_{s_1}^{s_2}R(s)
  \left(\frac{K(s)}s-P_n(s)\right)ds}\nonumber\\&&
  +\frac{4\alpha^2}\pi\Bigg[\frac1{2\pi i}\oint_{|s|=s_1}\Pi^{\rm had}(s)
  P_n(s)ds-\frac1{2\pi i}\oint_{|s|=s_2}\Pi^{\rm had}(s)P_n(s)ds\Bigg].
\end{eqnarray}

\section{Application of the method to $a_\mu^{\rm had}$}
The usual procedure to evaluate Eq.~(\ref{disper}) is to use the experimental
cross section up to some high momentum transfer and to calculate the remaining
part from QCD using local duality.

Indeed, in the energy region up to $1.2\GeV$ we use the $e^+e^-$ cross section
data set from Refs.~\cite{EidelmanJegerlehner} and~\cite{Akhmetshin}. The
corresponding dispersion integral reads
\begin{equation}\label{BreitWigner}
(\Delta  a_\mu^{\rm had})_1=\frac{\alpha^2}{3\pi^2}
  \int_{4m_{\pi}^2}^{(1.2\GeV)^2} \left[R^{e^+e^-}(s)+R^{\rm BW}(s)\right]
  \frac{K(s)}sds.
\end{equation}
Narrow resonances are added in explicit form. In Eq.~(\ref{BreitWigner}) the
missing resonances are parametrized by a Breit-Wigner fit $R^{\rm BW}(s)$ and
added to the remaining data sets.

In the subsequent interval from $1.2$ to $3.0\GeV$ the relevant $e^+e^-$ data
from Refs.~\cite{EidelmanJegerlehner}, combined with data sets from the BES
Collaboration~\cite{BES}, can be efficiently replaced by theoretical input
from QCD. By subtracting a polynomial function from the weight function
$K(s)/s$ we reduce the experimental input in this energy interval. Since the
polynomial function is an analytic function in the whole complex plane, the
remaining contribution can be evaluated off the real axis on a circular
contour in the complex plane using perturbative and nonperturbative QCD
results,
\begin{eqnarray}\label{split}
(a_\mu^{\rm had})_2&=&\frac{\alpha^2}{3\pi^2}
  \int_{s_1=(1.2\GeV)^2}^{s_2=(3.0\GeV)^2}R^{e^+e^-}(s)
  \left(\frac{K(s)}{s}-P_n(s)\right)ds\nonumber\\&&
  -\frac{4\alpha^2}{\pi}\frac1{2\pi i}\oint_{|s_2|=(3.0\GeV)^2}
  \left[\Pi^{\rm P}(s)+\Pi^{\rm NP}(s)\right]P_n(s)ds\nonumber\\&&
  +\frac{4\alpha^2}{\pi}\frac1{2\pi i}\oint_{|s_1|=(1.2\GeV)^2}
  \left[\Pi^{\rm P}(s)+\Pi^{\rm NP}(s)\right]P_n(s)ds\nonumber\\[3pt]
  &=&(a_\mu^{\rm exp})_2+(a_\mu^{\rm the})_2
\end{eqnarray}
where $n_f=3$ is taken for this interval.

At first glance the polynomial degree $n$ seems to be arbitrary. However, if
$n$ is chosen too large, both the theoretical error from the strong coupling
$\alpha_s(M_Z)$ and from unknown nonperturbative contributions arising from
higher order condensates lead to larger uncertainties. 

Moreover, we have to ensure that the calculated value is stable with respect
of the variations of the lower polynomial degrees $n$ where the less known
contributions from condensates are negligible. This point will help us find an
optimal energy domain for our method.

An optimal choice for the polynomial degree should guarantee a good
approximation of the polynomial function to the weight function and thus
effectively replace the experimental data by theoretical input while reducing
the total uncertainty from this region to a minimum.

Above $3.0\GeV$ the $e^+e^-$ data in Ref.~\cite{EidelmanJegerlehner,BES} show
low uncertainties and can be integrated directly,
\begin{equation}
(a_\mu^{\rm had})_3=\frac{\alpha^2}{3\pi^2}
  \int_{s_2=(3.0\GeV)^2}^{(40\GeV)^2}\left[R^{e^+e^-}(s)+R^{\rm BW}(s)\right]
  \frac{K(s)}{s}ds.
\end{equation}
In the last step the high energy tail starting from $40\GeV$ can be calculated
from theory using local duality. This is reasonable since there are no
resonance contributions above $40\GeV$. In this region it is sufficient to use
the lowest order contribution given by
\begin{equation}
\imag\Pi^{\rm had}(s)=3\!\!\!\sum_{{\it flavours}\ f}\!\!\!Q_f^2
  \frac{\alpha}{12\pi}\sqrt{1-\frac{4m_f^2}s}\left(1+\frac{2m_f^2}s\right).
\end{equation}

\begin{table}\begin{center}
\begin{tabular}{|l||r|r|r|r|}\hline
  in units of $10^{-10}$&$n=1$&$n=2$&$n=3$&$n=4$\\\hline\hline
$(a_\mu^{\rm the})_2$
  &$78.52$&$74.59$&$71.02$&$68.37$\\
  \hline
$(a_\mu^{\rm exp})_2$
  &$-8.63$&$-3.06$&$-0.93$&$-0.51$\\
  \hline\hline
$(a_\mu^{\rm had})_2$
  &$71.60$&$71.49$&$69.21$&$67.08$\\
  \hline\hline
data contribution&$12.1\%$&$4.3\%$&$1.3\%$&$0.8\%$\\\hline
\end{tabular}\end{center}
\caption{\label{tab1}Contributions to $(a_\mu^{\rm had})_2$ for
different fitting polynomial functions $P_n(s)$ with degree $n$. The purely
experimental and theoretical contributions according to Eq.~(\ref{split})
are listed separately. Because the error estimate becomes worse
for $n=3$ (cf.\ Table~\ref{tab2}), polynomial degrees $n=1$ and $n=2$ are
used to obtain a mean value and a methodical error estimate,
$(71.55\pm 0.13)\times 10^{-10}$.}
\end{table}

\section{Numerical results}
Using the recent $e^+e^-$ data from Refs.~\cite{EidelmanJegerlehner}
and~\cite{Akhmetshin} we obtain for the dominant low energy component of
$a_\mu^{\rm had}$,
\begin{equation}
(a_\mu^{\rm had})_1=(518.1\pm 4.8)\times 10^{-10}.
\end{equation}
We have added the statistical and point-to-point systematic errors in
quadrature. In addition to this experimental value we have to consider
contributions from the $\omega$ and $\phi$ resonances for which we obtain the
values $(38.9\pm1.4)\times 10^{-10}$ and $(40.4\pm1.3)\times 10^{-10}$,
respectively, from an integration over the Breit--Wigner distribution function
where the values for the parameters are taken from Ref.~\cite{PDG}.

\begin{figure}[ht]
\begin{center}
\epsfig{figure=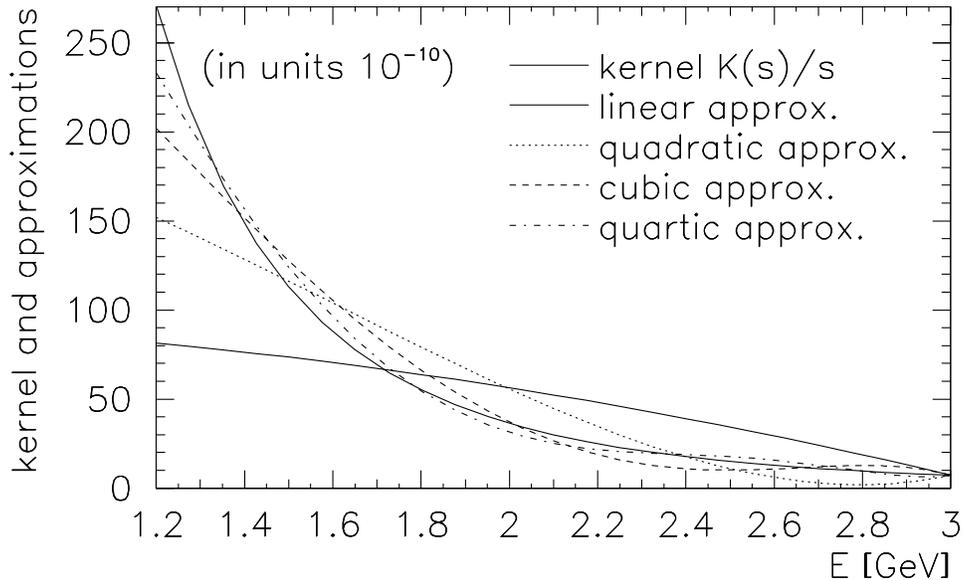, scale=0.8}
\end{center}
\caption{\label{fig2}The QED kernel $K(s)/s$ with fitting polynomial
  functions $P_n(s)$ of different degrees $n$. The polynomial functions were
  calculated by performing a least squares fit with no further constraints.}
\end{figure}

Applying our polynomial technique for the range between $1.2$ and $3.0\GeV$,
we incorporate QCD-expressions for the hadronic current-current correlator
with corresponding uncertainties for the parameters occuring in these
expressions. The errors from the masses of the light quarks $u$, $d$, and $s$
can be neglected at energies above $1\GeV$. Thus the theoretical uncertainty
is dominated by the strong coupling $\alpha_s(M_Z)$ whose error originates
from the uncertainty in the QCD scale
$\Lambda_{\overline{\rm MS}}=(380\pm 60)\MeV$. For the condensate terms we
assign generous errors of $100\%$. Thus we have
\begin{equation}
\langle\frac{\alpha_s}\pi G^2\rangle=(0.04\pm 0.04)\GeV^4,\qquad
\alpha_s|\langle\bar qq\rangle|^2=(4\pm 4)\times 10^{-4}\GeV^6.
\end{equation}
With increasing polynomial degrees the condensate error contributions grow
fast.
\begin{table}\begin{center}
\begin{tabular}{|l||r|r|r|r|}\hline
in units of $10^{-10}$&$n=1$&$n=2$&$n=3$&$n=4$\\\hline\hline
uncert.\ due to $\alpha_s(M_Z)$
  &$2.5$&$5.1$&$7.7$&$10.5$\\\hline
uncert.\ due to $\langle(\alpha_s/\pi)G^2\rangle$
  &$0.044$&$0.196$&$0.512$&$1.043$\\\hline\hline
uncertainty of $(a_\mu^{\rm the})_2$
  &$2.5$&$5.1$&$7.7$&$10.6$\\\hline
\end{tabular}\end{center}
\caption{\label{tab2}Theoretical uncertainties of $(a_\mu^{\rm had})_2$
from QCD parameters and condensates. Because the uncertainty is worse for
higher polynomial degrees, the degrees $n=1$ and $n=2$ are selected for the
analysis. The error estimate is given by the average value
$3.80\times 10^{-10}$. The uncertainty due to
$\alpha_s|\langle\bar qq\rangle|^2$ is less than $10^{-14}$ in all cases and
is therefore omitted in the table.}
\end{table}
However, as can be clearly seen in Tab.~\ref{tab1} or in Fig.~\ref{fig2},
good approximations to the kernel of the dispersion integral and consequently
a very high reduction of the experimental data influence in the interval
between $1.2$ and $3.0\GeV$ can be achieved even by low degrees $n=1$ and
$n=2$. This can be understood by the fact that one is far away from the
singularity at $s=0$ and thus the weight function can be well approximated by
low degree polynomials.

\begin{figure}[ht]
\begin{center}
\epsfig{figure=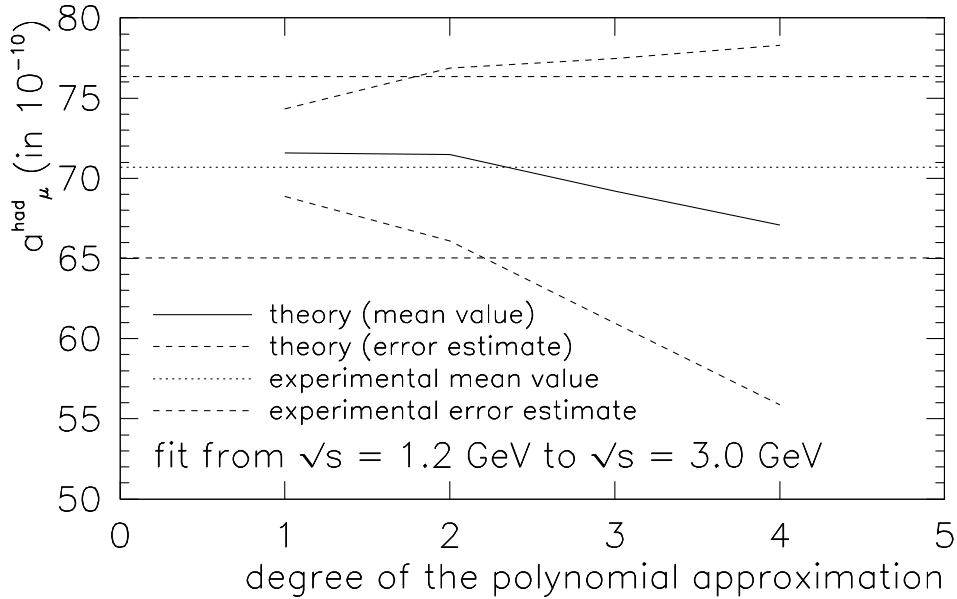, scale=0.8}
\end{center}
\caption{\label{fig3}Comparison of the l.h.s. and the r.h.s. of the sum rule
  Eq.~\ref{split} in the interval between $1.2\GeV$ and $3.0\GeV$. Dotted
  horizontal line: value of integrating the l.h.s. using experimental data
  including error bars. The points give the values of the r.h.s. integration
  for various orders $n$ of the polynomial approximation. Straight line
  interpolations between the points are for illustration only. The dashed
  lines indicate the error of our calculation.}
\end{figure}

We thus take the values $(a_\mu^{\rm had})_2$ for $n=1$ and $n=2$ and
calculate the algebraic mean. Therefore, we have to take into account a
small methodical error due to the variation of the polynomial degree,
\begin{equation}
(a_\mu^{\rm meth})_2=\pm0.06\times 10^{-10}.
\end{equation}
Summing up the error contributions in quadrature and the algebraic mean
uncertainty (for $n=1$ and $n=2$) from Table~\ref{tab2}, one obtains
\begin{equation}
(a_\mu^{\rm had})_2=(71.55\pm3.83)\times 10^{-10}.
\end{equation}

In the interval following $3.0\GeV$ up to the end of the $e^+e^-$ data set at
$40\GeV$ as given in Ref.~\cite{EidelmanJegerlehner} we obtain
\begin{equation}
(a_\mu^{\rm had})_3=(23.2\pm0.2)\times 10^{-10}.
\end{equation}
The contributions from narrow charmonium and bottonium resonances,
$(8.8\pm0.6)\times 10^{-10}$ and $(0.11\pm0.01)\times 10^{-10}$, respectively,
are taken by integrating over the narrow resonance distribution function where
values for the parameters are again taken from Ref.~\cite{PDG}.

We finally obtain a tiny continuum contribution from the last interval,
\begin{equation}
(a_\mu^{\rm had})_4=\frac{4\alpha^2}\pi\int_{(40\GeV)^2}^\infty
  \imag\Pi^{\rm had}(s)\frac{K(s)}sds=0.15\times 10^{-10}.
\end{equation}

\begin{table}\begin{center}
\begin{tabular}{|l|r|l|}\hline
interval for $\sqrt s$&contributions to $a_\mu^{\rm had}$&comments\\\hline
$[0.28\GeV,1.2\GeV]$&$(518.1\pm4.8)\times 10^{-10}$
  &$e^+e^-$ cross section data\\
$\omega$ resonance&$(38.9\pm1.4)\times 10^{-10}$&Breit-Wigner\\
$\phi$ resonances&$(40.4\pm1.3)\times 10^{-10}$&narrow resonances\\
$[1.2\GeV,3.0\GeV]$&$(71.6\pm3.8)\times 10^{-10}$&polynomial method\\
$J/\psi$ resonances&$(8.8\pm0.6)\times 10^{-10}$&narrow resonances\\
$[3.0\GeV,40\GeV]$&$(23.2\pm0.2)\times 10^{-10}$&$e^+e^-$ annihilation data\\
$\Upsilon$ resonances&$(0.11\pm0.01)\times 10^{-10}$&narrow resonances\\
$[40\GeV,\infty]$&$0.15\times 10^{-10}$&theory\\
top quark contr.&$<10^{-13}$&theory\\\hline
hadronic contr.&$(701.3\pm6.4)\times 10^{-10}$&\\\hline
\end{tabular}\end{center}
\caption{\label{tab3}The different contributions to the hadronic part of the
anomalous magnetic moment $a_\mu^{\rm had}$ of the muon.}
\end{table}
Summing up the contributions from the different energy intervals as shown in
Table~\ref{tab3}, we finally obtain
\begin{equation}
a_\mu^{\rm had}=(701.3\pm6.4)\times 10^{-10}
\end{equation}
for the hadronic contribution to $a_\mu$. We have thereby assumed that errors
with different origin are uncorrelated and have to be added quadratically in
order to obtain the total error.

\section{Higher order correction} 
\begin{figure}
\epsfig{figure=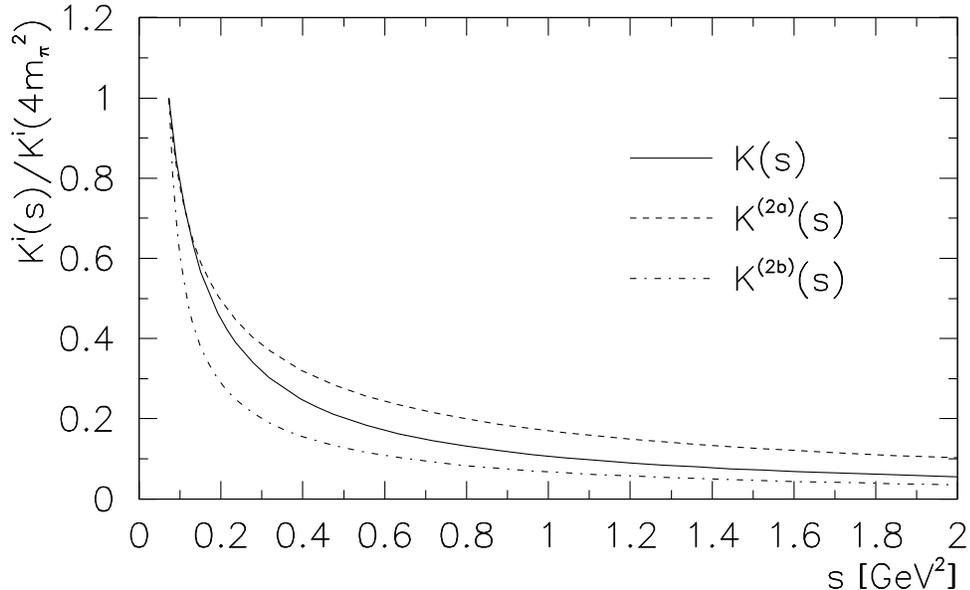, scale=0.8}
\caption{\label{fig4} Comparison of the LO and NLO kernels $K(s)$ and
  $K^{(2a,2b)}(s)$ normalized at $s=4m_\pi^2$. The functional behaviour of the
  three weight functions is quite similar.}
\end{figure}

In the next-to-leading order (NLO) we have to consider three types of diagrams,
those of type (2a) with an additional photon exchange, those of type (2b) with
an electron loop inserted in one of the photon lines in Fig.~\ref{fig1}, and
finally the one of type (2c) with two correlator functions included in the
photon line. For the former two the contribution reads
\begin{equation}
\label{exp0vacNLO}
a_\mu^{\rm had}({\rm NLO})=\frac13\pfrac{\alpha}{\pi}^3\int_{4m_\pi^2}^\infty
\frac{ds}{s}K^{(2)}(s)R(s).
\end{equation}
For numerical purposes it is convenient to represent the kernel functions
$K^{(2a)}(s)$ and $K^{(2b)}(s)$ in terms of power series expansions in terms
of $m^2/s$~\cite{Krause} ($m=m_\mu=105.6583568\pm0.0000052\MeV$~\cite{PDG} is
the mass of the muon). One has
\begin{eqnarray}
\lefteqn{K^{(2a)}(s)\ =\ 2\frac{m^2}{s}\Bigg\{
\left(\frac{223}{54}-\frac{\pi^2}{3}-\frac{23}{36}\ln\pfrac{s}{m^2}\right)}
  \nonumber\\&&
  +\frac{m^2}{s}\left(\frac{8785}{1152}-\frac{37\pi^2}{48}
  -\frac{367}{216}\ln\pfrac{s}{m^2}+\frac{19}{144}\ln^2\pfrac{s}{m^2}\right)
  \\&&
  +\frac{m^4}{s^2}\left(\frac{13072841}{432000}-\frac{883\pi^2}{240}
-\frac{10079}{3600}\ln\pfrac{s}{m^2}+\frac{141}{80}\ln^2\pfrac{s}{m^2}\right)
+\ldots\Bigg\},\nonumber\\
\lefteqn{K^{(2b)}(s)\ =\ 2\frac{m^2}{s}\Bigg\{
\left(-\frac{1}{18}+\frac{1}{9}\ln\pfrac{s}{m_f^2}\right)}
  \nonumber\\&&
  +\frac{m^2}{s}\left(-\frac{55}{48}+\frac{\pi^2}{18}
  +\frac{5}{9}\ln\pfrac{s}{m_f^2}+\frac{5}{36}\ln\pfrac{m^2}{m_f^2}
  -\frac{1}{6}\ln^2\pfrac{s}{m_f^2}+\frac{1}{6}\ln^2\pfrac{m^2}{m_f^2}\right)
  \\&&
  +\frac{m^4}{s^2}\left(-\frac{11299}{1800}+\frac{\pi^2}{3}
  +\frac{10}{3}\ln\pfrac{s}{m_f^2}-\frac{1}{10}\ln\pfrac{m^2}{m_f^2}
  -\ln^2\pfrac{s}{m_f^2}+\ln^2\pfrac{m^2}{m_f^2}\right)
+\ldots\Bigg\}\nonumber
\end{eqnarray}
where for $m_f$ we used the electron mass,
$m_f=510.998902\pm0.000021\keV$~\cite{PDG}, since the contributions from the
muon is already included in the contribution (2a). The contribution of the
$\tau$ is suppressed by $m_\tau^2/m_\mu^2$ and numerically negligible. The
kernels of the different orders in the electromagnetic coupling, $K(s)$ on the
one hand and $K^{(2a)}(s)$ and $K^{(2b)}(s)$ on the other hand, are compared
in Fig.~\ref{fig4}. For illustration, we like to sum the results from the
evaluation of Eq.~(\ref{exp0vacNLO}) which was performed in a completely
analogous way as in the previous chapter. The final result
$(-21.07\pm0.21)\times 10^{-10}$ in case of $K^{(2a)}(s)$ (cf.\
Table~\ref{tab4}) and $(10.78\pm0.08)\times 10^{-10}$ in case of $K^{(2b)}(s)$
(Table~\ref{tab5}) are in good agreement with the literature~\cite{Alemany}.
Finally, for the small contribution from diagrams of type (2c) with two
correlator functions included into the photon line we take the value
$(0.27\pm 0.01)\times 10^{-10}$ from Ref.~\cite{Krause}. We have checked that
to the required accuracy a recalculation using the method presented in this
paper is not necessary. The total hadronic NLO contribution then sums up to a
value of $(-10.3\pm0.2)\times 10^{-10}$, as compared to the value
$(-10.1\pm0.6)\times 10^{-10}$ given in Ref.~\cite{Krause}. Our central value
is $3\%$ smaller than that given in Ref.~\cite{Krause} and the error is
reduced from $0.6\times 10^{-10}$ to $0.2\times 10^{-10}$. Finally, summing up
leading and next-to-leading order contributions together with the
afore-mentioned QED and weak contributions and the so-called light-by-light
contribution $a_\mu^{\rm had}({\rm LBL})=(8.6\pm 3.5)\times 10^{-10}$
\cite{Knecht:2001qf,Kinoshita,Bijnens}, we obtain $a_\mu=(11\,659\,185.6\pm
6.4_{\rm had}\pm3.5_{\rm LBL}\pm0.4_{\rm QED+EW})\times 10^{-10}$.

\begin{table}\begin{center}
\begin{tabular}{|l|r|l|}\hline
interval for $\sqrt s$&contributions to $a_\mu^{\rm had}$&comments\\\hline
$[0.28\GeV,1.2\GeV]$&$(-14.04\pm0.12)\times 10^{-10}$
  &$e^+e^-$ cross section data\\
$\omega$ resonance&$(-1.11\pm0.04)\times 10^{-10}$&Breit-Wigner\\
$\phi$ resonances&$(-1.29\pm0.04)\times 10^{-10}$&narrow resonances\\
$[1.2\GeV,3.0\GeV]$&$(-2.84\pm0.14)\times 10^{-10}$&polynomial method\\
$J/\psi$ resonances&$(-0.44\pm0.03)\times 10^{-10}$&narrow resonances\\
$[3.0\GeV,40\GeV]$&$(-1.35\pm0.09)\times 10^{-10}$&$e^+e^-$ annihilation data\\
$\Upsilon$ resonances&$<10^{-13}$&narrow resonances\\
$[40\GeV,\infty]$&$<10^{-12}$&theory\\
top quark contr.&$<10^{-14}$&theory\\\hline
hadronic contr.&$(-21.07\pm0.21)\times 10^{-10}$&\\\hline
\end{tabular}\end{center}
\caption{\label{tab4}The different contributions to the hadronic part of the
anomalous magnetic moment $a_\mu^{\rm had}$ of the muon for the kernel
$K^{(2a)}$.}
\end{table}

\begin{table}\begin{center}
\begin{tabular}{|l|r|l|}\hline
interval for $\sqrt s$&contributions to $a_\mu^{\rm had}$&comments\\\hline
$[0.28\GeV,1.2\GeV]$&$(7.98\pm0.07)\times 10^{-10}$
  &$e^+e^-$ cross section data\\
$\omega$ resonance&$(0.59\pm0.02)\times 10^{-10}$&Breit-Wigner\\
$\phi$ resonances&$(0.62\pm0.02)\times 10^{-10}$&narrow resonances\\
$[1.2\GeV,3.0\GeV]$&$(1.09\pm0.01)\times 10^{-10}$&polynomial method\\
$J/\psi$ resonances&$(0.14\pm0.01)\times 10^{-10}$&narrow resonances\\
$[3.0\GeV,40\GeV]$&$(0.364\pm0.003)\times 10^{-10}$
  &$e^+e^-$ annihilation data\\
$\Upsilon$ resonances&$<10^{-13}$&narrow resonances\\
$[40\GeV,\infty]$&$<10^{-12}$&theory\\
top quark contr.&$<10^{-14}$&theory\\\hline
hadronic contr.&$(10.78\pm0.08)\times 10^{-10}$&\\\hline
\end{tabular}\end{center}
\caption{\label{tab5}The different contributions to the hadronic part of the
anomalous magnetic moment $a_\mu^{\rm had}$ of the muon for the kernel
$K^{(2b)}$.}
\end{table}

\section{Conclusion}
We have presented an alternative determination of the hadronic contribution to
the anomalous magnetic moment of the muon where we have made use of theoretical
QCD results to reduce the influence of the poor experimental data in the range
between $1.2$ and $3.0\GeV$. Our analysis includes the leading and
next-to-leading contribution.

Using the polynomial method in the range between $1.2$ and $3.0\GeV$, we could
suppress the influence of experimental data effectively and thereby reduce the
error on the determination of $a_\mu^{\rm had}$. Note that using this method
we were able to use a value for the lower limit of this range lower than
usually used for pure QCD methods.

Using only $e^+e^-$ data and QCD as input, we obtain a result which is
$1.6\sigma$ away from the measured world average, stating that the deviation
between theory and experiment may be smaller than commonly suggested. Other
recent results to compare with are $(11\,659\,180.9\pm7.2_{\rm had}\pm
3.5_{\rm LBL}\pm0.4_{\rm QED+EW})\times10^{-10}$~\cite{Davier:2003pw} and
$(11\,659\,166.9\pm7.4)\times10^{-10}$~\cite{Hagiwara}.

\subsection*{Acknowledgements}
We would like to thank K.~Schilcher for discussions. S.G. acknowledges a grant
given by the DFG, Germany through the Graduiertenkolleg ``Eichtheorien'' at
the University of Mainz.

\end{document}